\documentstyle[aps,epsfig,preprint]{revtex}
\tightenlines
\begin{document}
\def\be{\begin{equation}}
\def\en{\end{equation}}
\def\bq{\begin{eqnarray}}
\def\eq{\end{eqnarray}}
\def\noi{\noindent}
\def\bi{\bigskip}
\def\tr{{\rm tr}}
\def\ii{\'{\i}}
\def\l({\left(}
\def\r){\right)}

\title{$U_A(1)$ Symmetry Breaking and  the Scalar Sector of QCD}
\author{ M. Napsuciale and S. Rodriguez \\
{\small\it Instituto de Fisica, Universidad de Guanajuato,\\
AP E-143, 37150, Leon, Guanajuato, Mexico }}
\maketitle

\begin{abstract}
It is shown that most of the unusual properties of the lowest lying scalar 
(and pseudoscalar) mesons can be understood, at the qualitative and 
quantitative level, on the 
basis of the breakdown of the $U_A(1)$ symmetry  coupled to the 
vacuum expectation values of scalars by the spontaneous breaking of chiral 
symmetry.
\end{abstract}

\section{Introduction}

\bi
The light quark sector of QCD acquires a $U(N_f)_L\otimes U(N_f)_R$ 
symmetry and scale  invariance in the limit of  $N_f$ massless quarks. 
This limit is a reasonable starting point for constructing  
effective theories. 
The $u$ and $d$ quark masses are small enough compared to the naive 
confinement scale ($\Lambda_{QCD}$) to make the procedure reliable. 
Incorporating the $s$ quark \cite{chpt} makes sense 
once one realizes that the relevant scale is 
$\Lambda_{\chi}\approx 4\pi f_\pi$  \cite{manohar}. 
The conventional approach to low energy QCD is to assume that the 
octet of lowest lying pseudoscalar mesons $\{ \pi,~K,~\eta\}$ 
are approximate Goldstone Bosons (GB) associated with the spontaneous 
breakdown of the $SU(3)_L\otimes SU(3)_R$  symmetry.
This belief is supported by the smallness of $\pi$, $K$ and $\eta$ 
meson masses as compared to the typical hadronic scale of $ 1 GeV$. 
The approach leads to a so far successful description of
pseudoscalar meson physics where the relevant QCD Green 
functions are systematically expanded in powers of 
$\partial / \Lambda_\chi$, ~ $m_q/\Lambda_\chi  $, where 
$\Lambda_\chi$ stands for the chiral symmetry breaking scale, in  the 
so-called Chiral Perturbation Theory ($\chi PT$) expansion 
\cite{chpt}. 
The remaining pseudoscalar, the $\eta ^\prime$ meson, has a large mass 
which has been related to the breakdown of the $U_A(1)$ symmetry  
\cite{thooft,thooftu2,witten}.

The symmetries of massless QCD are reduced to smaller symmetries in many ways. 
Both scale invariance and $U_A(1)$ symmetry are broken at the quantum level 
\cite{shifman}. The residual $SU(3)_L\otimes SU(3)_R\otimes U(1)_V$ symmetry 
is spontaneously reduced down to $SU(3)_V$ (the vector $U(1)$ symmetry being 
trivial for mesons). Finally, 
if one introduces mass terms, $SU(3)_V$ reduces down to isospin, or 
completely breaks down, depending on the choice for the quark masses. 

The most important contribution to  the breakdown of the $U(1)_A$ symmetry
comes from Euclidean classical field 
configurations with non-trivial 
topology (instantons). These effects are modulated 
by the $e^{-8\pi^2/g^2_s}$ factor, where $g_s$ denotes the strong coupling constant,
thus being  small at 
high energies but  becoming important  at low energies \cite {thooft}.

Physics beyond the low energy region (where $\chi PT$ is strictly valid) 
and up to the $\phi(1020)$ mass requires to introduce 
all the relevant degrees of freedom, namely, the $\eta^\prime $ meson, the 
well established $a_0(980)$ and $f_0(980)$ scalar mesons and the lowest lying 
vector meson nonet.
The formulation of effective theories including the $\eta '$ and its 
relation to the $U(1)_A$ anomaly was  firstly done in 
\cite{veneziano,schechter,witten1}.  In some of these pioneering works 
\cite{veneziano,schechter} the starting point is a 
Lagrangian which exhibits chiral symmetry realized in a linear way. 
This amounts to consider scalar fields as the chiral partners of 
pseudoscalars. In particular in \cite{schechter} the relation to the 
instanton induced quark interaction discovered by 
't Hooft is established by expanding one of the terms in the 
Lagrangian. Scalar fields are 
then integrated out and authors focus on the properties of the vacuum and 
 on the pseudoscalar phenomenology. 

Over the past few years the problem of the scalar mesons has been under 
intense debate. The identification of $\bar q q$ scalar mesons is still 
awaiting for a final answer. This problem is related to the unique QCD 
prediction for the existence of bound states of gluons (glueballs) and 
the possibility for the existence of hybrid states. The well established 
low mass scalar mesons are: the $f_0(980)$, $f_0(1370)$ and the recently 
resurrected  $f_0(400-1200)$ (or $\sigma$) in the  $I=0$ sector; the 
$a_0(980)$ and $a_0(1450)$ on the isovector side, and the isospinor 
$K^*_0(1430)$  \cite {pdg}. 
In addition to these states there are claims for the existence of signals 
of other scalar mesons below 1 GeV \cite{sigma,kappa,schechter1,napsu,tornq}. 
In particular, the approach summarized in \cite{schechter1} which is  based 
on a non-linear chiral Lagrangian including scalar degrees of freedom 
concludes the existence of a light isoscalar scalar $\sigma$ and an 
isovector scalar $\kappa$ with a mass around 850 MeV. This is consistent 
with results in \cite{napsu} though these approaches differ in 
the mixing of scalars and its interpretation. The differences come from the 
fact that $U_A(1)$ anomaly effects are not taken into account in 
\cite{schechter1}, while they are considered as fundamental in 
\cite{napsu,tornq,thooft1}. As we shall see below the picture arising in the 
later case is completely consistent and most of the controversial properties 
of scalar mesons find a natural and simple explanation 
in this framework.  

Many other schemes have been put forth trying to understand the lowest lying 
scalar meson properties. A $\bar qq$ structure for the $a_0(980)$ and 
$f_0(980)$ was put on doubt since the corresponding  quark models were not 
able to explain the tiny coupling of these mesons to two photons 
\cite{barnes} and the quark content suggested by the 
mass spectrum was incompatible with the main decay channels 
\cite{jaffe}. More complicated models such as four 
quark states \cite{jaffe} and molecules \cite{barnes} were advocated to 
explain these points. This is a very active field and no definitive 
conclusion has been reached  as to which states are to be 
considered as $\bar qq$, multi-quark, molecule, gluonia or  hybrid states 
\cite{narison}.  

In this work we intend to shed some light on this problem. To this end 
let us first exhibit  some of the properties of scalar  
(and pseudoscalar) mesons which seem to be in conflict with a $\bar q q$ 
interpretation.

On the base of a naive constituent quark model we  expect the
$^1S_0$ $\bar q q$ states ($q=u,d$) to have a mass 
$\simeq 2m_q\approx 630~ $MeV where $m_q$ 
stands for  $u$ or $d$ constituent quark mass, which we consider as equal and 
evaluate to one third of the 
proton mass. In a similar way the $\bar q s$ ($\bar s s$) states should have 
a mass $\approx 870~ MeV$ ( $\approx 1010$~MeV) if we evaluate the 
constituent strange quark mass to $m_s\simeq m_\Omega /3 \approx 555~$MeV.
In this picture,
a $^3S_1$ $\bar q q$ state would be slightly heavier than the $^1S_0$ state 
and a $^3P_0$ $\bar q q$ state should have an even larger mass. 

The lowest lying vector mesons fulfill these naive expectations, it is in 
this  sense a ``well-behaved'' sector. The pseudoscalar and scalar sectors, 
however, are not. The pions and  
kaons are lighter than expected from naive quark model considerations. 
The smallness of the pion and kaon masses  
is qualitatively understood on the base of spontaneous breaking of $\chi$-ral 
symmetry. However, even with this explanation in mind, there exist at least 
one problem in this sector which is usually overlooked and which has to do 
with the quark content of pseudoscalars. Consider the 
isoscalar-pseudoscalar states with well defined quark content 
(flavor states): 
$\eta_{ns} \equiv (\bar uu +\bar dd)/\sqrt{2}$ and $\eta_{s} \equiv \bar ss $.
Since they are not the physical states, they  must be mixed somehow to yield 
the physical  $\eta$ and $\eta^\prime$ mesons. Independently of the 
mechanism which mixes the flavor isoscalar-pseudoscalar states, a 
straightforward analysis yields the relations
\be
m^2_{\eta_{ns}}= m^2_{\eta} \cos^2\phi_P +  m^2_{\eta^\prime} \sin^2\phi_P ,~~~
m^2_{\eta_{s}}= m^2_{\eta} \sin^2\phi_P +  m^2_{\eta^\prime} \cos^2\phi_P, 
\label{etans}
\en
\noi where the information on the mixing mechanism is hidden in the
mixing angle in the flavor basis $\phi_P$. This angle has been 
estimated as $\phi_P\approx 39.5^\circ $ \cite{feldmann}, corresponding 
to an angle $\theta_P\approx -15.2^\circ $ in the naive single-angle 
description of pseudoscalar mesons mixing in the usual singlet-octet basis. 
Introducing this information in Eq.(\ref{etans}) we obtain 
$m_{\eta_{ns}} \approx 741 ~MeV$ and $m_{\eta_s} \approx 816 ~MeV$.
The obvious question here is: Why is the $\eta_{ns}$ ( a $\bar q q$ state 
differing from the pion only in isospin quantum numbers)  much heavier 
than the pion, even more massive than  the $\bar qs$ ($K$)  
and as heavy as the (purely strange) $\eta_s$ state ?

Concerning the scalar sector things are much more involved since even an 
undoubtful identification of the corresponding nonet is still missing. Assume 
that the $a_0(980)$ and $f_0(980)$ mesons are the $I=1$ and $I=0$ 
members of the $\bar q q$ scalar nonet. Their nearby degeneracy suggest that 
they are the scalar mesons analogous to the $\rho(770)$ and $\omega(780)$ 
vector mesons which are also almost degenerate in mass. 
On this identification grounds the $f_0(980)$ should be predominantly 
non-strange but then: How can one explain its strong coupling to $\bar K K$?. 
This could be qualitatively understood if the $f_0(980)$ had a strong strange 
component but then one does not understand why this meson is  almost 
degenerate in mass with the non-strange $a_0(980)$.
On the other hand the $a_0(1450)$ and the $f_0(1350)$ seem to be heavy enough 
to be out of the scope of the  naive quark model considerations. The same can 
be said for the vector $K^*_0(1430)$.
Another possibility is the  existence of  light 
scalar mesons: $\sigma(400-600)$ and $\kappa (\approx 900)$. It seems that we 
have now a general consensus on the existence of a broad scalar structure in 
the low energy region \cite{sigma} though it is not clear yet 
what its nature is and which mechanism makes this meson so light.
The $\kappa (900)$ is a  more controversial object \cite{kappa} and its 
existence has not been firmly established. 

As mentioned above, the vector mesons spectrum  is in qualitative  agreement 
with naive quark model expectations. Thus, whatever  the explanation  for the
 scalar and pseudoscalar mesons spectrum may be, it should have no effect on 
the vector sector.
 
In this letter we suggest that all the points raised above can be 
understood by considering the instanton induced quark interaction which is 
also responsible for the $U_A(1)$ symmetry breaking.
The appealing property of this interaction is that it does not affect the 
vector sector since vector mesons consist of quark-anti-quark pairs that 
are either both left-handed or both right-handed and do not match the 
quantum numbers of the instanton induced interaction \cite{thooft,thooft1}. 
This, and the fact that it is able to explain why the $\eta^\prime$ is not 
a Goldstone Boson make this interaction a natural candidate for explaining 
the unusual properties of pseudoscalar mesons and their chiral partners, 
the scalar mesons. Thus, we require a model considering both nonets in a 
chiral symmetric way and which takes into account  
t'Hooft interaction.

\section{The Model}

A model with the characteristics mentioned at the end of the previous 
section has been reconsidered recently \cite{napsu,tornq,thooft1}.
The model, incorporates pseudoscalar and scalar degrees of freedom, 
exhibits chiral symmetry and incorporates $U_A(1)$ breaking in a 
phenomenological way. A $U(2)\otimes U(2)$ version of this model was 
formulated in connection with the explanation of the $U_A(1)$ anomaly 
\cite{thooftu2}. Amazingly, similar models were put forward even before 
the birth of QCD \cite{gasio}.
The Lagrangian is given by
\be
{\cal L}= {\cal L}_{sym}  +{\cal L}_{SB} \label{lagrangian}
\en
where ${\cal L}_{sym}$ denotes the $U(3)\times U(3)$ symmetric Lagrangian:
\be
{\cal L}_{sym}=\bigl<\frac{1}{2}\l(\partial_\mu M\r)
 \l(\partial^\mu M^{\dagger}\r) \bigr>  -\frac{\mu^2}{2} 
X \l(\sigma ,P\r)-\frac{\lambda}{4} Y\l(\sigma ,P\r)-
\frac{\lambda^\prime}{4} X^2 \l(\sigma , P\r) \label{symlag}
\en
\noi with $M=\sigma + i P$, and $X,Y$ stand for the $U(3)\times U(3)$ 
chirally symmetric terms:
\be
 X\l(\sigma ,P\r)=\bigl< M M^{\dagger} \bigr> , ~~~
 Y\l(\sigma,P\r)= \bigl<(M M^{\dagger})^2\bigr>.
\en
We closely follow the conventions in Ref.\cite{napsu}. In particular, we use 
$F\equiv \frac{1}{\sqrt{2}} \lambda_i f_i$ with $F=\sigma,~P$; 
$f_i=\sigma_i,~ p_i$; $i= 0..8$ and $\lambda_i$ denote Gell-Mann matrices. 
{}For further details the interested reader is referred to Ref.\cite{napsu}.
Chiral and $U_A(1)$ symmetries are  explicitly broken by 
\be
{\cal L}_{SB} =  \bigl< c\sigma \bigr> - \beta Z\l(\sigma,P\r) \label{asymlag}
\en
where $c\equiv \frac{1}{\sqrt{2}} \lambda_i c_i$, with $c_i$ constant and
\be
Z\l(\sigma ,P\r) =  \mbox{det} \l(M\r)+ \mbox{det} \l(M^{\dagger}\r) 
\en
The most general form of $c$ which preserves isospin and gives PCAC is such 
that the only non-vanishing coefficients are $c_0$ and $c_8$. The former 
gives, by hand, the pseudo-scalar nonet a common mass, while the later 
breaks the $SU(3)$ symmetry down to isospin. These parameters can be  
related to quark masses in QCD.

The $Z$ term in Eq.(\ref{asymlag}) corresponds to the  instanton induced quark 
interaction in the case when the instanton angle $\theta_{inst} $ is aligned to 
zero. This interaction has the form of a determinant in flavor space and breaks 
$U(3)_L\otimes U(3)_R$ into $SU(3)_L\otimes SU(3)_R \otimes U(1)_V $ \cite{thooft1}. 

The  linear $\sigma$  term in Eq. (\ref{asymlag}) induces $\sigma$-vacuum transitions 
which give to  $\sigma$  fields  a  non-zero vacuum expectation value 
(hereafter denoted by $\{~\}$). Linear terms can be eliminated  from  the  
theory  by  performing a shift to a new 
scalar  field  $S=\sigma -V$ such that $\{S\}=0$,  where  $V\equiv\{\sigma\}$.
This shift generates new three-meson interactions and mass terms.    

\noi
{}For the sake of simplicity let us write $V$=Diag$ (a,a,b)$ where $a,b$ are 
related to $\{\sigma \}$ through
\be
a=\frac{1}{\sqrt{3}}\{\sigma_0\}+\frac{1}{\sqrt{6}}\{\sigma_8 \}, ~~~~
b=\frac{1}{\sqrt{3}}\{\sigma_0\}-\frac{2}{\sqrt{6}}\{\sigma_8\}
\en
Meson masses generated by this procedure are:

i) Non-mixed sectors.
\bq \nonumber
m^2_\pi =\xi+2\beta b+\lambda a^2, &\quad&
m^2_K =\xi+2\beta a+\lambda (a^2-ab+b^2), \label{nonmixed} \\
m^2_a = \xi-2\beta b +3\lambda a^2,   &\quad&
m^2_\kappa =\xi-2\beta a+\lambda (a^2+ab+b^2),  
\eq

\noi where $a$ and $\kappa$ denote the scalar mesons analogous to
 $\pi$ and $K$ respectively  and we used the shorthand notation 
$\xi = \mu^2 +\lambda '(2a^2+b^2)$.
 
ii) Mixed sectors.
\bq
{\cal L}^P_2 &=&-\frac{1}{2} \l(m^2_{0P} P^2_0+ m^2_{8P} P^2_8 + 
2m^2_{08P} P_0 P_8\r),  \label{mixlag} \\
{\cal L}^S_2 &=&-\frac{1}{2} \l(m^2_{0S} S^2_0+ m^2_{8S} S^2_8 + 
2m^2_{08S} S_0 S_8\r) \nonumber
\eq
\noi where

\bq \nonumber
m^2_{8P} &=& \xi+ {1\over 3} [\lambda \l(a^2 +2 b^2\r) + 2\beta \l(4a - b\r)], \\
 \nonumber
m^2_{0P} &=& \xi+ {1\over 3} [\lambda \l(2 a^2 + b^2\r) - 4 
\beta \l(2a+b\r)], \\ \nonumber
m^2_{08P} &=& \frac{\sqrt{2}}{3} \l(a-b\r)[\lambda \l(a+b\r) + 2\beta ],
\label{mixed1} \\
m^2_{8S} &=& \xi+\frac{1}{3} [-2\beta \l(4a-b\r)+3\lambda \l(a^2+2b^2\r)+
4\lambda^\prime \l(a-b\r)^2],  \\ \nonumber
m^2_{0S} &=&\xi+\frac{1}{3} [4\beta\l(2a+b\r)+3\lambda \l(2a^2+b^2\r)+
2\lambda^\prime \l(2a+b\r)^2], \\ \nonumber
m^2_{08S}&=&\frac{\sqrt{2}}{3}\l(a-b\r)[-2\beta+3\lambda\l(a+b\r)+
2\lambda^\prime\l(2a+b\r)].\\ \nonumber
\eq
As we are interested in the quark content of fields it is convenient to analyze 
the mixed sector in the flavor ($\{|s>, |ns>\}$) basis \cite{mike} defined by:
\be
 \eta_{ns}=\sqrt{1\over 3} P_8  + \sqrt{2\over 3}P_0, ~~~~~~~ 
 \eta_{s} =-\sqrt{2\over 3 } P_8  + \sqrt{1\over 3} P_0 
\en 
\noi with analogous relations for the scalar ($\sigma_{ns},\sigma_s$) mixed fields.
In this representation, the mass terms in the Lagrangian read 
\bq
{\cal L}^P_2 =-\frac{1}{2} \l(m^2_{\eta_{ns}} \eta^2_{ns}+ m^2_{\eta_s} \eta^2_s + 
2m^2_{\eta_{s-ns}} \eta_{ns} \eta_s\r)  \label{mixlagsns} \\
{\cal L}^S_2 =-\frac{1}{2} \l(m^2_{\sigma_{ns}} \sigma^2_{ns}+ m^2_{\sigma_s} 
\sigma^2_s + 2m^2_{\sigma_{s-ns}} \sigma_{ns} \sigma_s\r) \nonumber
\eq
\noi where
\bq
\begin{array}{lclclcl}
m^2_{\eta_{ns}} &=&\xi-2\beta b + \lambda a^2 , & &
m^2_{\sigma_{ns}} &=&\xi+2\beta b + 3\lambda a^2 +4\lambda^\prime a^2 , \\
m^2_{\eta_s} &=&\xi + \lambda b^2 , & &
m^2_{\sigma_s} &=&\xi + 3\lambda b^2 + 2\lambda^\prime b^2, \\
m^2_{\eta_{s-ns}} &=& -2\sqrt{2}\beta a , & &   
m^2_{\sigma_{s-ns}} &=& 2\sqrt{2}\l(\beta+\lambda^\prime b\r)a .
\end{array}
\label{mixed2}
\eq
Many of our initial concerns can be qualitatively understood from Eqs. 
(\ref{nonmixed},\ref{mixed1},\ref{mixed2}).
The first point to be noticed is that, in this model,
 {\it the $U_A(1)$ anomaly gets coupled to the v.e.v.'s of scalar 
fields by the spontaneous 
breaking of chiral symmetry and contributes, via this effect, to 
the masses of all fields entering the theory except the strange fields} . 
In the remaining of the paper we will call this the ``anomaly-vacuum'' 
effect. The mass terms arising as a consequence of this effect, have 
exactly the same origin as the mass terms coming from the $\phi^4$ terms 
in the $SU(2)\times SU(2)$ Linear Sigma Model of Gell-Mann and Levy. 
The novelty here is the existence of the three-meson determinantal 
instanton induced interaction which couples either two pseudoscalars to 
one scalar or three scalars (other possibilities being forbidden by parity). 
In the case when one of the scalars in the vertex has the same 
quantum numbers as the vacuum (the isospin singlets 
${1\over \sqrt{2}}(\bar uu + \bar dd)$ or $\bar ss$ ) a mass term is 
generated by replacing this leg by the corresponding vacuum expectation value. 
In particular the vertex with a strange pseudoscalar a non-strange 
pseudoscalar and an isosinglet scalar field (which the determinantal 
structure dictates to be the non-strange isosinglet) gives rise to a mass 
term proportional to the strength of the anomaly, which  mixes the strange 
and non-strange pseudoscalar fields as shown in Eq.(ref{mixed2}. 
Similar phenomena occur in the scalar sector. 
The strength of the 
mixing due to the anomaly is exactly of the same size but opposite sign in 
the pseudoscalar and scalar sectors. In the latter case,  
there exist an additional contribution coming from the $\phi^4$ interaction 
in the Lagrangian  Eq.~(\ref{symlag}) whose strength is measured by 
$\lambda^\prime$.

The extraction of the parameters of the model
has been done in \cite{napsu,tornq,thooft1} using well 
known information on the pseudoscalar sector. 
In this concern, it is worth mentioning that the outcome 
of the model strongly depends on the input used. In Ref.~\cite{napsu} 
 pseudoscalar meson masses ($ m_\pi,~m_\eta,~m_{\eta '},~ m_K$) and the 
pion decay constant ($f_\pi$) were used in order to 
fix the parameters entering the model ($\xi, \lambda, \beta 
,a, x=(b-a)/2a $). Ref. \cite{tornq} uses 
$m_\pi,m_K,~m^2_\eta+m^2_{\eta^\prime},~f_K$ and $f_\pi$ as input while Ref. 
\cite{thooft1} uses the pseudoscalar mixing angle in the singlet-octet basis 
$\theta_P$ and $m_\pi,m_K,~m_\eta,~m_{\eta '}$. The outcome is different 
in all these three cases. In particular, the last approach yields a heavy 
scalar nonet. It must be pointed out, however, that according to recent 
work \cite{leutwyler}, the proper description of pseudoscalar mixing in the 
singlet-octet basis requires two mixing angles. 
In this concern, the use of the strange-non-strange basis is more 
appropriate since, in this basis, pseudoscalar mixing can be described using a 
single angle \cite{feldmann}. As we shall see below, the pseudoscalar 
spectrum is consistent, within the model, with a small mixing angle in the 
singlet-octet basis. An explanation for the physics behind this quantity 
requires the improvement of the model.
The mixing  angle of pseudoscalars in the flavor basis($\phi_P$) and its 
scalar analogous ($\phi_S$) can be extracted from (\ref{mixlagsns}) by 
diagonalizing the Lagrangian. A straightforward calculation yields
\be
\sin2\phi_P = {2 m^2_{\eta_{s-ns}} \over m^2_{\eta^\prime} - m^2_{\eta} },~~
\sin2\phi_S = {2 m^2_{\sigma_{s-ns}} \over  
m^2_{\sigma^\prime} - m^2_{\sigma} }\label{pseudmix}
\en
\noi where $\sigma^\prime $ and $\sigma $ denote the physical 
isoscalar-scalar mesons. Alternatively, 
\be
\cos2\phi_P = { m^2_{\eta_{s}}- m^2_{\eta_{ns}} \over m^2_{\eta^\prime} - 
m^2_{\eta} },~~
\cos2\phi_S = {m^2_{\sigma_{s}} - m^2_{\sigma_{ns}} \over 
m^2_{\sigma^\prime} - m^2_{\sigma} }.\label{pseudmix1}
\en
The important point is that the parameter which measures the strength of the 
$U_A(1)$ breaking, $\beta$, turns out to be negative with the conventions in 
Eq.(\ref{asymlag}): $\beta\approx -1.5 ~GeV$. The value of $a$ can be 
fixed from $a=f_{\pi}/\sqrt{2}=65.7~ MeV $, whereas the value of $b$, which we 
rewrite in terms of $x=(b-a)/2a$ can be extracted 
by the procedure used in \cite{tornq} which yields $x=0.22$
or from the input in  \cite{napsu} which gives $x=0.39$. The 
values of $\lambda$ and $\lambda^\prime$ are positive. Their actual values 
depend on the input used but in general $\lambda^\prime $  turns out to be 
small \cite{napsu,tornq}. Using the approach of Ref. \cite{napsu}, 
Eqs. (\ref{pseudmix},\ref{pseudmix1}) yield 
\be
\sin 2\phi_P = 0.9202 ~~~ \cos 2\phi_P=-0.3911,
\en
which imply a value  $\phi_P\approx 56.7^\circ $ for the pseudoscalar 
mixing angle in the $s-ns$ basis. 
 This corresponds to a small angle 
in the singlet-octet basis ($\theta_P = \phi_P - 54.7^\circ \approx 
+2^\circ $) consistent with results 
in \cite{tornq} ($\theta\approx -5^\circ$)
\footnote{In Ref. \cite{napsu} the pseudoscalar mixing angle  
in the $s-ns$ basis was 
extracted from the sine relation Eq.(\ref{pseudmix}) and  estimated as 
$\phi_P \approx 33.3^\circ$. A 
careful analysis of the diagonalization process 
shows that the correct value arises from the cosine relation and its actual 
value is $\phi_P \approx  56.7^\circ $.
 The problem when using the  sine 
relation is that it 
does not distinguish between $\phi_P$ and ${\pi \over 2} - \phi_P$ which is 
also a solution. We appreciate illuminating correspondence with  
Prof. G. 't Hooft which helped to clarify the input scheme dependence and 
lead us to reconsider the extraction of the pseudoscalar mixing angle.}. 
The difference in these approaches comes from the use of 
$f_K$ as input in \cite{tornq}, instead of the combination 
$m^2_{\eta^\prime} -m^2_{\eta}$ used in \cite{napsu}. 

The extraction of the scalar mixing angle requires to fix the coupling 
$\lambda^\prime$ which enters the pseudoscalar and the unmixed scalar 
sectors only in the combination 
$\xi = \mu^2 +\lambda ^\prime (2a^2+b^2)$ as can be seen from  
Eqs.(\ref{nonmixed},\ref{mixed2}). Thus, fixing this parameter necessarily  
requires to use information on the mixed scalar sector. In \cite{napsu} the 
masses of physical scalar mesons where studied as a function of this 
parameter. This analysis leads to the identification of the 
isoscalar-scalar mesons with the $\sigma(400-600)$ and $f_0(980)$. 
This identification yield $\lambda^\prime \approx 4$ which predicts a 
scalar mixing angle $\phi_S \approx -14^\circ $. This result is 
consistent with the analysis of $f_0\to \gamma\gamma $ \cite{luna} and 
the recently measured $\phi\to\pi^0\pi^0\gamma $ \cite{novosibirsk,nalu}. 
The small scalar mixing angle is consistent with  a 
mostly $\bar ss$ structure for the $f_0(980)$ meson and a nearby 
$(\bar uu +\bar dd)/\sqrt{2}$ for the sigma meson. The isovector and 
isospinor scalar fields are identified with the $a_0(980)$ and 
$\kappa (\approx 900)$ in the model ( the 
procedure followed in \cite{tornq} identifies $\kappa$ with 
$K^*_0(1430)$, this is a consequence of using different input and to the 
high dependence of the outcome on the  input scheme mentioned above). 

\section{Pseudoscalar spectrum}

The purpose of this letter is to remark the role played by the 
$U_A(1)$ anomaly in structuring  the mass spectrum of scalar 
(and pseudoscalar) mesons. To this end, let us qualitatively analyze 
the splittings  arising from from Eqs. (\ref{nonmixed},\ref{mixed2}) 
in order to disentangle the different contributions to the meson masses. 
In this concern, it is important to recall that the extraction of the 
parameters yields a negative sign for $\beta$ in Eq.(\ref{asymlag}) which 
quantify the strength of the anomaly.
{}From Eqs.(\ref{nonmixed},\ref{mixed2}) we obtain the relations
\bq
m^2_{\eta_{ns}}-m^2_\pi &=& -4\beta b,  \\
m^2_{\eta_{ns}}-m^2_K &=& -2\beta \l(a+b\r) - \lambda b\l(b-a\r), \label{splitetans}
\eq
\noi which reveal that, in the absence of the anomaly, the pion and 
non-strange eta are degenerate and the kaon is heavier than both of them, 
which is consistent with the expectations of naive quark models.
Thus, the $\eta_{ns}-\pi$ splitting is due to the ``anomaly-vacuum''
 effect which pushes the pion and kaon masses down and the non-strange 
eta mass up as can be seen from Eqs. (\ref{nonmixed},\ref{mixed2}). 
As to the  $\eta_{ns}-K$ splitting there is an additional contribution 
coming from  the invariant in (\ref{symlag}) whose strength is measured by the 
$\lambda $ coupling constant. From now on we will call such kind of 
contributions as the ``normal'' effects. They are ``normal'' in the sense 
that, in the absence of the anomaly, the splittings between pseudoscalar 
mesons are proportional to this coupling constant times the SU(3) symmetry 
breaking factor $(b-a)$ which can be related to the difference of strange 
and non-strange quark masses. In the case at hand, the ``normal'' effect 
tends to make  $K$ heavier than $\eta_{ns}$, but  the ``anomaly-vacuum''
effect goes in the opposite direction. The latter effect is stronger than 
the former thus rendering the non-strange eta heavier than the kaon. 

Similar results are obtained for the $\eta_s-\pi$ and $\eta_s-K$ splittings 
as can be seen from the following relations 
\bq
m^2_{\eta_s}-m^2_\pi&=& -2\beta b + \lambda \l(b+a\r)\l(b-a\r) \\
m^2_{\eta_s}-m^2_K&=& -2\beta a + \lambda a\l(b-a\r). \label{splitetas}
\eq
\noi In this case, both effects interfere constructively, reinforcing the 
corresponding splittings. The individual effects of the anomaly can be read 
from Eqs.(\ref{nonmixed},\ref{mixed2}). The anomaly leaves the strange eta 
untouched while, as noticed above pushes the pion and kaon masses down.

The next interesting effect in the pseudoscalar spectrum has to do with the 
$K-\pi$ splitting. From the relation
\be
m^2_K-m^2_\pi = \l(-2\beta+\lambda b\r)\l(b-a\r), \label{kapi}
\en
\noi we see that $K-\pi$ mass splitting is proportional to the $SU(3)$ 
symmetry  breaking. It must be noticed , however, that the ``normal'' effect 
is enlarged  by the ``anomaly-vacuum'' effect  thus making pions much 
lighter than kaons.
Finally the $\eta_s-\eta_{ns}$ splitting is 
\be
m^2_{\eta_s}-m^2_{\eta_{ns}}= 2\beta b +\lambda \l(b+a\r)\l(b-a\r). \label{splitetasns}
\en
\noi In the absence of the anomaly the normal pattern appears, i.e. the 
strange field is heavier than the non-strange field. Turning on the anomaly 
modifies this pattern. The anomaly does not affect the strange pseudoscalar 
mass but it does push up the non-strange pseudoscalar.  If we consider the 
mixing angle predicted by the model, the later turns out to be 
heavier than the former.

\section{Scalar spectrum}

Let us now turn to the scalar sector. The first point to be emphasized is 
that, the ``anomaly-vacuum'' contribution  systematically has the opposite 
sign in the pseudoscalar and scalar sectors ( we will call this the 
``sign''effect ). The reader can be easily convinced of this by looking at 
the mass relations in   Eqs.(\ref{nonmixed},\ref{mixed2}). There are two 
important consequences of the ``sign'' effect. The first one is that the 
$a_0$ and $\kappa$ masses are pushed {\it up} by the anomaly 
(in contrast to the pseudoscalar analogous which are pushed 
{\it down}) and the non-strange sigma is pushed {\it down} 
(unlike the non-strange pseudoscalar which is pushed {\it up}) . 
The second effect has to do with the mixing of strange and non-strange 
isoscalar scalar fields. The ``anomaly-vacuum'' effect in the mixing of 
scalars is exactly of the same size as in pseudoscalars but with the 
opposite sign . In this case, there exists an additional contribution 
coming from one of the chiral invariants in the Lagrangian, 
whose strength is measured by the $\lambda^\prime$ coupling. As discussed 
in \cite{tornq} this term corresponds to disconnected quark diagrams, 
thus being suppressed by the Okubo-Zweig-Izuka rule. This term, however, 
becomes  relevant to the mixing of scalars since it has the opposite sign 
to the anomaly and interferes destructively with it, thus rendering the 
scalars less strongly mixed than pseudoscalars.  
 
The $\sigma_{ns}-a$ and $\sigma_{ns}-\kappa$ splittings are given by
\bq
m^2_{\sigma_{ns}}-m^2_a &=&4\beta b + 4\lambda^\prime a^2 \\
m^2_{\sigma_{ns}}-m^2_\kappa &=& 2 \beta \l(a+b\r) - \lambda \l(2a+b\r)\l(b-a\r) + 
 4 \lambda^\prime a^2. \nonumber
\eq
\noi Notice that in the absence of the anomaly and the OZI forbidden 
$\lambda^\prime $ coupling the $a$ and $\sigma$ fields have the same 
mass  and the $\kappa$  meson is slightly heavier as expected from 
their constituent quark content. 
The individual effects of the anomaly mentioned above  (the anomaly pushes 
the sigma mass down and the $a$ and kappa masses up) causes the $a-\sigma $ 
splitting making the $a$ field heavy and the sigma field light. In other 
words, {\it the $\sigma$ meson is light as a consequence of the 
``anomaly-vacuum'' and the ``sign'' effects.} 
In the case of the $\kappa - \sigma$ splitting, these  mesons are pushed 
in opposite directions by the anomaly thus reinforcing the 
``normal'' pattern. In addition we have to 
take into account $\lambda^\prime$ contribution in both cases.

The corresponding relations for $\sigma_s$ are
\bq
m^2_{\sigma_{s}}-m^2_a &=& 2\beta b +3\lambda \l(a+b\r)\l(b-a\r) 
+ 2 \lambda^\prime b^2 \\
m^2_{\sigma_{s}}-m^2_\kappa &=& 2\beta a +\lambda \l(a+2b\r)\l(b-a\r)+
2\lambda^\prime b^2. \nonumber
\eq
\noi Turning off the anomaly (and the $\lambda^\prime$ contribution ) 
yields a pattern as expected from the naive constituent quark picture. 
The strange scalar is heavier than the non-strange one and the kappa meson. 
It worth noticing that the strange scalar field 
is not affected by the ``anomaly-vacuum'' effect (see Eqs.(\ref{mixed2})), 
but the chiral partner of the pion, the $a$ field, is pushed up by the 
anomaly rendering it almost degenerate with the physical (mostly strange) 
$f_0$ meson. Thus {\it the $a_0 - f_0$ degeneracy is accidental 
and has its origin in the instanton induced quark interaction}. 
In both equations above, the  ``normal'' effect and the 
$\lambda^\prime$ contributions to the splittings have the same sign. These 
``normal'' splittings are 
 largely canceled by the 
``anomaly-vacuum''effect. 
This explains the $a_0-f_0$ degeneracy and the close 
value of the $\kappa$ meson mass. 

The $a_0-\kappa$ splitting has the same 
structure as the corresponding pseudoscalar relation in 
Eq.~(\ref{kapi}), namely
\be
m^2_\kappa -m^2_a = \l(2\beta + \lambda \l(2a+b\r)\r)\l(b-a\r)
\en
but in this case the individual``anomaly-vacuum'' effects go in the opposite 
direction due to the ``sign''effect. Both meson masses are pushed up by the 
anomaly with different strengths as the anomaly is coupled to 
different v.e.v.'s . The result is a not so strong anomaly effect in the 
splitting which nevertheless is strong enough to largely cancel 
the ``normal'' splittings due to quark masses making these mesons roughly 
equally heavy.

Finally, the $\sigma_s -\sigma_{ns}$ splitting can be read from 
Eqs.(\ref{mixed2}) as 
\be
m^2_{\sigma_s}-m^2_{\sigma_{ns}} = -2\beta b + 3 \lambda \l(b+a\r)\l(b-a\r) + 
2\lambda^\prime \l(b^2- 2a^2\r).
\en
\noi In case of a vanishing anomaly, we obtain the ``normal''pattern 
again. Switching on the anomaly affect the non-strange $\sigma$ mass only 
( see Eq.(\ref{mixed2} ))
making this meson  light. In other words,  
the ``anomaly-vacuum'' and ``normal''contributions 
interfere constructively in the splitting. This makes 
the non-strange $\sigma$ much lighter than the strange sigma. The 
naively expected pattern is reinforced by the  ``anomaly-vacuum'' effect 
in contrast to the pseudoscalar analogous case 
in Eq.(\ref{splitetasns}) where the anomaly overcome
the ``normal'' effect. 

The qualitative  analysis above can be graphically described as  in Figs. 1-2, 
where the splittings due to the ``anomaly-vacuum''  and  $SU(3)$ 
breaking effects are shown.
\bi
\bi

\vskip2ex
\centerline{
\epsfxsize=450 pt
\epsfbox{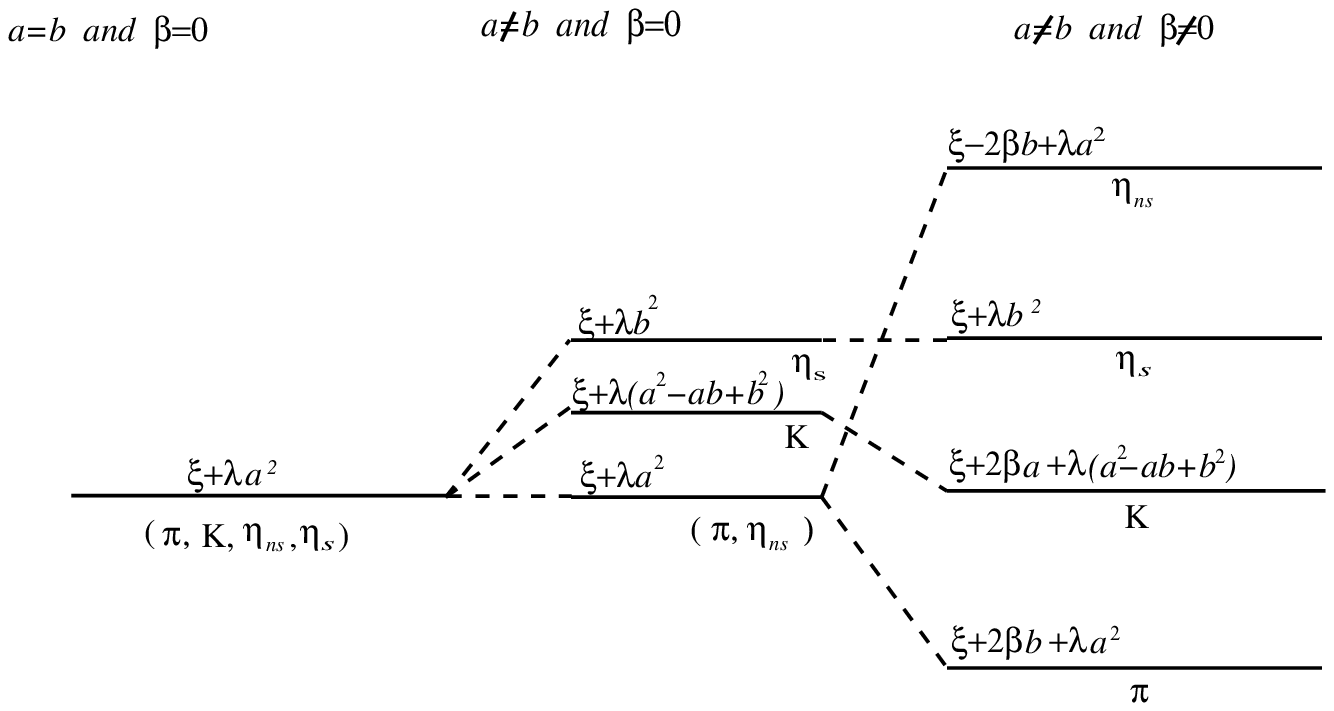}}
\begin{center}
{\small{Fig.1.} Effects of  $U_A(1)$ symmetry breaking in the 
pseudoscalar sector. }
\end{center}

\vskip2ex
\centerline{
\epsfxsize=400 pt
\epsfbox{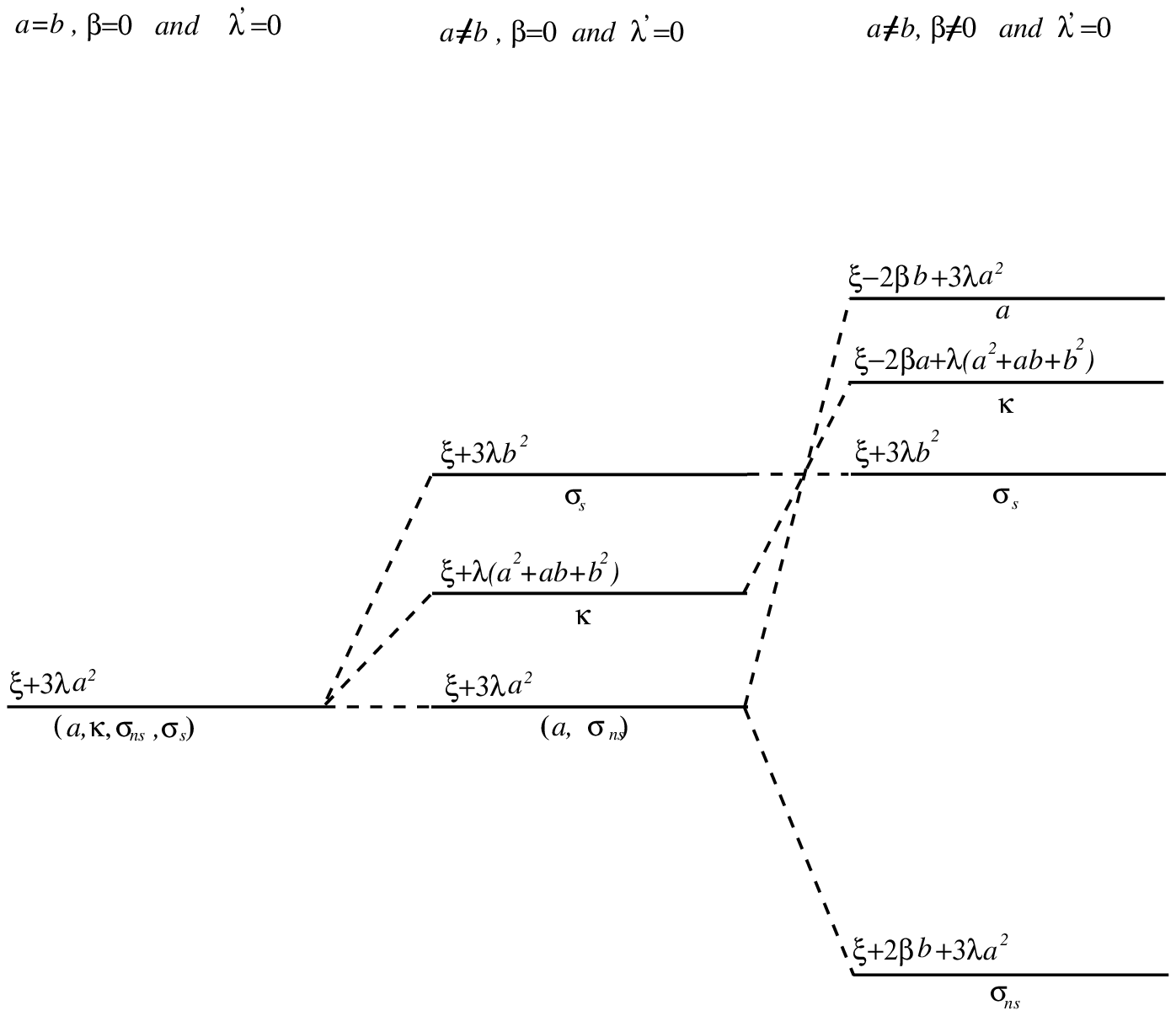}}
\begin{center}
{\small{Fig.2.} Effects of  $U_A(1)$ symmetry breaking in the scalar sector. }
\end{center}


 

\section{ Summary }

Summarizing, we have obtained a qualitative (and quantitative) 
understanding of the  lowest lying scalar (and pseudoscalar)  
meson spectrum on the basis of a 
simple model  which incorporates  the most relevant 
properties of QCD in the corresponding energy region, namely, the appropriate 
degrees of freedom, spontaneous breaking of chiral symmetry and $U(1)_A$ 
symmetry breaking. Most of  the unusual 
properties of  both sectors are explained by the coupling of the 
$U(1)_A$ anomaly to the vacuum expectation values of scalars by the 
spontaneous breaking of chiral symmetry (``anomaly-vacuum'' effect ) 
which supplies mesons with not yet considered mass terms.

Certainly this is only a first step in the elucidation of the quark 
structure of scalar mesons, and we have considered the $U_A(1)$ anomaly 
as the only striking effect (beyond chiral symmetry breaking).
It still remain to conciliate this point of view with the 1/N perspective  
\cite{shuryak} which is a main task.
The fact that so many phenomena are explained by a single effect makes 
this model attractive as a starting point for more elaborate and 
systematic analysis. It is also reassuring that quantitative estimates 
of the model  predictions for the scalar meson decays where intermediate 
vector mesons can not appear or their contribution is known to be small 
are successful. Recently, it has been shown that the puzzling 
$a_0(980)\to\gamma\gamma$ and $f_0(980)\to \gamma\gamma$ decays are properly 
described within the model \cite{luna} and the recent measurements for 
the $\phi \to \pi^0\pi^0\gamma$ decay \cite{novosibirsk} are consistent with 
model predictions \cite{nalu}. 

The improvement of the model is necessary in order to reach 
a more complete description of meson properties. 
This is shown by its failure to satisfactorily describe the pseudoscalar 
mixing angle. Other effects such as the hadronic loops considered 
in \cite{lipkin} could remedy this.  
Another possibility is to include degrees of freedom which 
could be relevant to the physics in this energy region and so far have not 
been considered. In particular, glueballs degrees of freedom could have 
some effects via their mixing with pure quarkonium states. A step in this 
direction is the work in \cite{nagy} where glueball degrees of freedom 
are considered in  a framework close to the one used here and its worthy to 
analyze its consequences in detail.
A different line of action which is worth exploring is the 
relation of the instanton induced quark interaction with the recently raised 
group theoretical arguments in favor of 
the unitary symmetry as an accidental symmetry due to the 
gluon anomaly \cite{mariana}.


\begin{thebibliography}{99}

\bibitem{chpt} S. Weinberg,
Physica {\bf A96}, 327 (1979); 
J. Gasser and L. Leutwyler, Nucl. Phys. {\bf B250}, 465 (1985);
J. Gasser, M.E. Sainio and V. Svarc,
Nucl. Phys. {\bf B307}, 779 (1988). 
{}For a review see e.g.``Dynamics of the Standard 
Model'' J.F. Donoghue, E. Golowich, and B.R. Holstein 
(Cambridge Univ. Press, 1992).

\bibitem{manohar} A. Manohar and H. Georgi, Nucl. Phys.  {\bf B234}, 
189 (1984).  

\bibitem{thooft} G. 't Hooft, Phys. Rev. Lett {\bf 37} (1976); 
Phys. Rev. {\bf D14}, 3432 (1976); 
{\it Err.} Phys. Rev. {\bf D18}, 2199 (1978). 

\bibitem{thooftu2} G. 't Hooft, Phys. Rep. {\bf 142}, 357, (1986).
\bibitem{witten} E. Witten, Nucl. Phys. {\bf B156}, 269 (1979); 
                G. Veneziano, Nucl. Phys. {\bf B159}, 213 (1979).
\bibitem{shifman} M. A. Shifman, Phys. Rep. {\bf 209}, 341 (1991).

\bibitem{veneziano} P. Di Vecchia and G. Veneziano, Nucl. Phys. 
{\bf B171}, 23 (1980).

\bibitem{schechter} C. Rozenzweig, J. Schechter, and C. G. Trahern, 
Phys. Rev. {\bf D21}, 3388 (1980).

\bibitem{witten1} E. Witten, Ann. Phys. {\bf 128}, 363 (1980).

\bibitem{pdg} D.E. Groom {\it et. al.}, Eur. Phys. Jour. {\bf C15}, 1 
(2000). 

\bibitem{sigma}  N. Tornqvist and M. Roos, Z. Phys. {\bf C68},
 647 (1995);
 N. Tornqvist and M. Roos, Phys. Rev. Lett. {\bf 76}, 1575 (1996);
 N. Tornqvist and M. Roos, Phys. Rev. Lett. {\bf 78}, 1604 (1997);
Phys. Rev. Lett. {\bf 78}, 1603 (1997); 
S. Ishida, Prog. Theor. Phys. {\bf 95}, 745 (1996);
M. Ishida, Prog. Theor. Phys. {\bf 96}, 853 (1996);
F. Sannino and J. Schechter, Phys. Rev. {\bf D52}, 96 (1995); 
M. Harada, F. Sannino and J. Schechter, Phys. Rev. {\bf D54}, 1991 (1996); 
Phys. Rev. Lett. {\bf 78}, 1603 (1997); 
J.A. Oller and E. Oset, Nucl.Phys. {\bf A629}, 739 (1998); 
Nucl.Phys. {\bf  A620}, 438 (1997);
Phys.Rev. {\bf D60}, 074023 (1999).

\bibitem{kappa} D. Black et. al. Phys. Rev. {\bf D58}, 054012 (1998); 
J. Oller, E. Oset and  J. R. Pelaez, Phys. Rev. {\bf D59}, 074001 (1999);
S. N. Cherry and M. Pennington; [hep-ph/0005208].  

\bibitem{schechter1} H. Fariborz, [hep-ph/9908329].

\bibitem{napsu} M. Napsuciale [hep-ph/9803396], unpublished.

\bibitem{tornq}  N. Tornqvist, Eur. Phys. Jour. {\bf C11}, 359 (1999). 

\bibitem{thooft1} G. 't Hooft, [hep-th/9903189]. 

\bibitem{barnes} J. Wenstein, N. Isgur Phys. Rev {\bf D27}, 588 (1983); 
T. Barnes, Phys. Lett {\bf B165}, 434 (1985).

\bibitem{jaffe} R.J. Jaffe, Phys. Rev. {\bf D15}, 267 (1977); 
N.N. Achasov and V. N. Ivanchenko, Nucl Phys. {\bf B315}, 465 (1989).

\bibitem{narison} S. Narison, Nucl. Phys. {\bf A675}, 54 (2000);
[ hep-ph/9909470]. 

\bibitem{gasio} M.Levy Nuov. Cim. {\bf LIIA}, 7247 (1967); 
S. Gasiorowicz and D.A. Geffen Rev. Mod. Phys. {\bf 41}, 531 (1969); 
J. Schechter and Y. Ueda Phys. Rev. {\bf D3}, 2874 (1971);
L.H. Chan and R.W. Haymaker Phys. Rev. {\bf D7}, 402 (1973); 
Phys. Rev. {\bf D7}, 415 (1973). 

\bibitem{mike} M. D. Scadron Phys. Rev. {\bf D26}, 239 (1982). 

\bibitem{leutwyler} G. M. Shore, Nucl. Phys. {\bf B569}, 107 (2000);
H. Leutwyler, Nucl. Phys. Proc. Suppl.{\bf 64}, 223 (1998); 
R. Escribano and J.M. Frere, Phys. Lett. {\bf B459}, 288 (1999); 
T. Feldmann, P. Kroll and  B. Stech, Phys. Rev. {\bf D58}, 114006 (1998).     

\bibitem{feldmann} T. Feldmann, Int. Jour. Mod. Phys. {\bf A15}, 159 (2000).

\bibitem{luna} J.L. Lucio and M. Napsuciale Phys. Lett. {\bf B454}, 365 (1999).

\bibitem{novosibirsk} R.R. Akhmetsin {\it et. al} Phys. Lett {\bf B462}, 
380 (1999).

\bibitem{nalu} J.L.Lucio and M. Napsuciale, to appear in Proceedings of the 
Workshop on Physics and Detectors for DAFNE  (DAFNE99), Frascati, Italy, 
Nov. (1999); [hep-ph/0001136].

\bibitem{shuryak} T. Schafer and E. Shuryak, hep-lat/0005025; 
N. Isgur and H. B. Thacker, hep-lat/0005006.

\bibitem{lipkin} H. Lipkin and B. Zou, Phys. Rev.
{\bf D53}, 6693 (1996); 
B. Zou, Phys. Atom. Nucl. {\bf 59}, 1427 (1996);[ hep-ph/9611283].

\bibitem{nagy} M. Nagy, M. K. Volkov and V. L. Yudichev, hep-ph/0003200.

\bibitem{mariana} M. Kirchbach, Phys. Rev. {\bf D58}, 117901 (1998);
                  M. Kirchbach, Phys. Lett. {\bf B455}, 259 (1999).               

\end{thebibliography}
\end{document}